\documentclass[11 pt]{article}

\usepackage{graphicx}
\usepackage{setspace}

\usepackage{amsmath}
\usepackage{amsfonts}
\usepackage[top=3cm, bottom=3cm, left=3cm, right=3cm]{geometry}
\begin{document}

\title{A Hybrid Local Search for \\Simplified Protein Structure Prediction}

\author{Swakkhar Shatabda, M.A. Hakim Newton, Duc Nghia Pham and Abdul Sattar \thanks{All the authors are affiliated with Queensland Research Laboratory, NICTA and Institute for Integrated and Intelligent Systems (IIIS), Griffith University}}
\date{}
\maketitle
%\keywords{Protein Structure Prediction, Discrete Lattices, Local Search, HP Energy Model, FCC Lattice, Heuristics.}

\abstract{Protein structure prediction based on Hydrophobic-Polar energy model essentially becomes searching for a conformation having a compact hydrophobic core at the center. The hydrophobic core minimizes the interaction energy between the amino acids of the given protein. Local search algorithms can quickly find very good conformations by moving repeatedly from the current solution to its ``best'' neighbor. However, once such a compact hydrophobic core is found, the search stagnates and spends enormous effort in quest of an alternative core. In this paper, we attempt to restructure segments of a conformation with such compact core. We select one large segment or a number of small segments and apply exhaustive local search. We also apply a mix of heuristics so that one heuristic can help escape local minima of another. We evaluated our algorithm by using Face Centered Cubic (FCC) Lattice on a set of standard benchmark proteins and obtain significantly better results than that of the state-of-the-art methods.}

\section{{Introduction}}
\label{secProtein}
  
Proteins are the most important organisms in a living cell. The function of a protein depends on the three dimensional native structure that it folds into in a particular environment. Knowledge about this native structure can have an enormous impact on the field of drug discovery. Computational methods for protein structure prediction (PSP) are of great interest since the \textit{in vitro} laboratory methods are very slow, expensive, and error-prone. In absence of any known templates for the proteins, computational methods like homology modeling and threading are not applicable. \textit{Ab initio} methods start from scratch and perform a search on the conformational space of structures. High resolution models require all atomic details and are not computationally preferable. Moreover, the contributing factors of different forces of the energy function are unknown and the space of the conformations is very large and complex. Simplified models, though lack many details, provide realistic backbone for the proteins. 

Even in the simplified models, the search space is not suitable for complete search methods. Local search methods can produce good quality conformations very quickly. However, they suffer from re-visitation and stagnation. The nature of the stagnation also depends on the fitness function. In Hydrophobic-Polar (HP) energy model, PSP essentially becomes searching for a conformation having a compact hydrophobic core at the center. Local search algorithms can quickly find a compact core. However, once such a core is found, the search stagnates and spends enormous effort in quest of an alternative core.  

In this paper, we attempt to restructure segments of a conformation with a very compact core. We select one large segment or a number of small segments and apply exhaustive local search. The total number of amino-acid positions affected by the segments selected in an iteration is dynamically adjusted with the stagnation period. We also use a tabu list to prevent recently changed amino-acid positions from being modified again. Moreover, we apply a mix of heuristics so that one heuristic can help escape local minima of another. These heuristics are derived from domain specific knowledge. Experimental results show that our approach significantly outperforms the state-of-the-art methods on a set of standard benchmark proteins on Face Centered Cubic (FCC) lattice.

\section{{Problem Definition}}
\label{secBack}
A protein is a polymer of amino-acids, which are also called monomers. There are only 20 different amino acids. In the simplified model, each amino acid is represented by the position of its $\alpha$-$C$ atom. The position is a valid point in the three dimensional lattice. Moreover, a simplified function is used in calculating the energy of a conformation. Note, every two consecutive monomers in the sequence are in \textit{contact} or neighbors on the lattice (called the \textit{chain constraint}) and two monomers can not occupy the same lattice point (called the \textit{self avoiding constraint}).

\subsection{FCC Lattice}
Face Centered Cubic (FCC) lattice is preferred to other lattices since it has the highest packing density and it provides the highest degree of freedom for placing an amino acid. Thus, FCC lattice provides a realistic discrete mapping for proteins. An FCC lattice has 12 basis vectors: $\vec{v_1}=(1,1,0)$, $\vec{v_2}=(-1,-1,0)$, $\vec{v_3}=(-1,1,0)$, $\vec{v_4}=(1,-1,0)$, $\vec{v_5}=(0,1,1)$, $\vec{v_6}=(0,1,-1)$, $\vec{v_7}=(0,-1,-1)$, $\vec{v_8}=(0,-1,1)$, $\vec{v_9}=(1,0,1)$, $\vec{v_{10}}=(-1,0,1)$, $\vec{v_{11}}=(-1,0,-1)$, $\vec{v_{12}}=(1,0,-1)$. Two lattice points $p,q$ $\in$ $\mathbb{L}$ are said to be in contact or $neighbors$ of each other, if $q = p+\vec{v_i}$ for some vector $\vec{v}_i$ in the basis of $\mathbb{L}$. 

\subsection{HP Energy Model}
The basic Hydrophobic-Polar (HP) model introduced in \cite{hpdill} divides the amino-acids into two groups: hydrophobic H and hydrophilic or polar P. The amino acid sequence of a given protein then becomes a string $s$ of the alphabet $\{H,P\}$. The free energy calculation for the HP model, shown in (\ref{eqHP}) counts only the energy interactions between two non-consecutive amino acid monomers.

\begin{equation}
 E=\sum_{i,j:i+1<j} c_{ij}.e_{ij}
\label{eqHP}
\end{equation}
Here, $c_{ij}$ = 1 when two monomers $i$ and $j$ are neighbors (or in contact) on the lattice and 0, otherwise. The other term, $e_{ij}$ is calculated depending on the type of amino acids: $e_{ij} = -1$, if $s_i = s_j = H$ and 0, otherwise. Note that, minimizing the summation in (\ref{eqHP}) is equivalent to maximizing the number of non-consecutive H-H contacts. 

 Using the HP energy model together with the FCC lattice, the simplified PSP problem is defined as: given a sequence $s$ of length $n$, find a self avoiding walk $p_1\cdots p_n$ on the lattice such that the energy defined by (\ref{eqHP}) is minimized.

\section {{Related Work}}
\label{secRel}
Various techniques and their hybridization have been applied to solve PSP. Genetic algorithms with Metropolis conditions as an acceptance criteria are found more efficient than Monte Carlo simulation \cite{Unger2}. Genetic algorithms are subsequently improved by other researchers \cite{geneticT,Rashid2012GAPlus}. The Constraint-based Hydrophobic Core construction (CHCC) algorithm \cite{yue1995forces} successfully produced optimal structures for the famous Tortilla benchmarks by using constraint programming techniques. Constraint-Based Protein Structure Prediction (CPSP) tools \cite{mann2008cpsp} were developed based on this CHCC algorithm. Later on, another constraint solver, COLA \cite{DalPalu2007COLA} was developed using several biologically inspired heuristics. It solved the problem with finite domains of the existing SICStus libraries. A two stage optimization method was proposed in \cite{twostage} to improve the solutions generated by CPSP tool by using simulated annealing in the second stage. Further, a large neighborhood search method \cite{UllahS10}, when run for a long time, produced better results than simulated annealing. In this work, constraint programming was used for neighborhood generation.

Tortilla benchmarks were solved for the first time in \cite{cebrian2008protein} by using FCC lattice and tabu meta-heuristics. In a subsequent work, more improved results were achieved by applying large neighborhood search and constraint programming \cite{dotu2008protein,dotu2011protein}. A memory based approach \cite{mem2012} on top of the local search framework \cite{dotu2011protein} further improved the results for these benchmarks and other larger proteins taken from CASP.

\section{{Our Approach}}
\label{secMem}

Local search methods produce good results quickly. In HP energy model, they form a compact core of hydrophobic residues at the center of the conformation and search can not progress unless the core is broken and an alternate core is formed \cite{mem2012}. Even when guided by a good heuristic, the search oscillates within the same region of the search space and fails to improve. This obvious nature of local search algorithms results in stagnation. Large neighborhood techniques are adopted to handle this situation in protein structure prediction \cite{UllahS10,dotu2011protein} and in other domains as well \cite {BentH07}. Most of these algorithms depend on constraint programming for neighborhood generation. In this paper, we propose a hybrid local search that can improve the solutions by restructuring a single or multiple segments of the selected points, and thus breaking the compact core to create an alternative core. Our algorithm belongs to local search family and do not use constraint programming for neighborhood generation. The pseudo-code of our method is given below:

\begin{footnotesize}
\begin{verbatim}
Procedure LWS(Protein seq)
1   initializeTabu()
2   while ++it <= maxIt do
3     selectSegmentType()
4     selectSegmentVariables()
5     generateMoves()
6     selectHeuristic()
7     simulateMoves()
8     selectBestMove()
9     executeSelectedMove() 
10    updateTabuList()
11    if not Improving for 
12         maxStable steps then
13      maxStable *= factor 
14      segmentSize++
15    end if
16  end while
17  return globalBestStructure
End Procedure.
\end{verbatim}
\end{footnotesize}

At each iteration, our algorithm selects a number of variables depending on the segment size and segment type (Lines 3-4). Then, all feasible moves are generated (Line 5) using the selected variables. These variables are essentially Cartesian co-ordinates of amino-acid residues. Once the feasible neighborhood is generated, it then simulates the moves and calculates the changes in the heuristic selected in line 6. The simulate function (Line 7) temporarily updates the selected heuristic in an incremental fashion. Once the best move is selected (Line 8), the conformation is updated by executing the move (Line 9). The execution of the move permanently updates the fitness functions and the heuristics. If there is a new global best, all the parameters are reset to the initial condition. Stagnation occurs if there is no improvement in the global best for $maxStable$ steps. The segment size is increased and the stagnation parameter, $maxStable$ is multiplied by a $factor$ at stagnation. We also maintain a tabu list that prevents selecting recently modified variables. 

\subsection{Algorithm Details}
In the rest of this section, we describe different parts of the algorithm in details.       

\begin{figure}[htb]
  \begin{center}
	\begin{footnotesize}
    \begin{tabular}{c}
          \includegraphics[width=.30\textwidth]{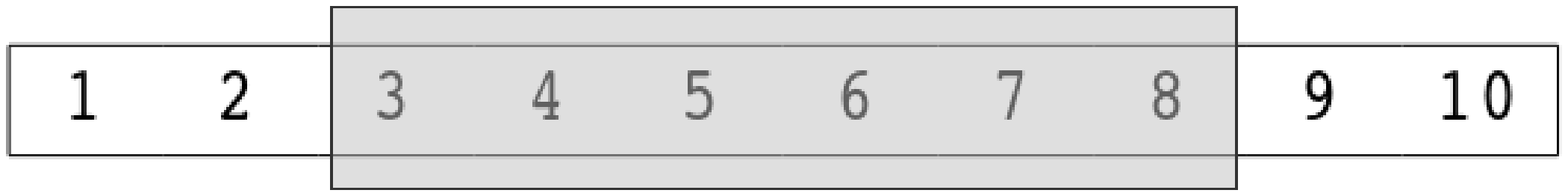}
      \\
	(a) Single Segment \\
          \includegraphics[width=.30\textwidth]{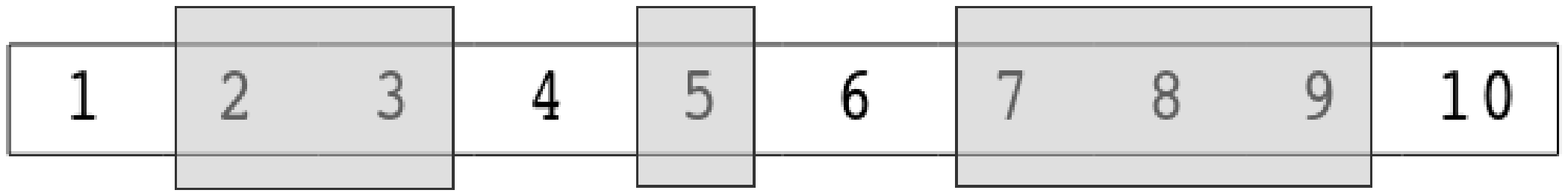}
      \\
      (b) Multiple Segments\\
    \end{tabular}
\end{footnotesize}
  \end{center}
  \caption{\footnotesize Two types of segments, for $segmentSize =6$ }%
  \label{figWin}%
\end{figure}

\paragraph{Segment Types.}
We select one large segment or a number of small segments (see Figure~\ref{figWin}). At each iteration, a number of variables are selected to fill these segments, which are then used in generating moves. The purpose of the segment search is to locally re-optimize the structure within the segment using exhaustive search. Large segment type allows re-structuring of a large subsequence of the protein while the multiple segment type re-optimizes multiple subsequences simultaneously. We select a segment type randomly at each iteration. The total number of amino-acid positions in the segments selected in an iteration is $segmentSize$.

\paragraph{Variable Selection.} We maintain a tabu list to prevent recent moves. The tabu tenure is selected randomly using a uniform distribution from the range $[4,sequenceLength/8]$. In case of the single large segment, we select the variables from the range $[-segmentSize/2$, $+segmentSize/2]$ around a randomly selected point, if none of them are in the tabu list. Though the tabu record is kept for single points, this mechanism requires all the points in a segment to be out of tabu list and hence a different part of the structure is guaranteed to be selected for re-structuring at each iteration. In the case of multiple segments, we randomly select the variables that are not in the tabu list. It creates multiple segments each containing points from different parts of the structure.

\paragraph{Segment Search.}
Once the variables are selected, the algorithm then generates all the possible moves for those variables, keeping the rest of the chain un-affected. Pseudo-code for the procedure $generateMoves()$ is given below:
\begin{footnotesize}
\begin{verbatim}
Procedure generateMoves(Conf c, Segment w)
1    generator = initialize()
2    do
3      for (all points p in w)
4        newPoint = getPosition(generator)
5        if(occupied(newPoint))
6          skip(generator, p)
7          exit for
8        end if
9        add newPoint to currentMove
10     end for
11     if(!skip)
12       add currentMove to moveList
13     end if 
14   while(next(generator))    
End Procedure
\end{verbatim}
\end{footnotesize}
The algorithm starts with an initial generator string that assigns the same direction vector to all the positions. Each direction vector is one of the basis vector between two consecutive points. For each of the points in the segment, a new point is calculated using the generator string (line 4). If that position is already occupied, then the rest of the generator string is ignored by calling the method $skip(genrator,p)$. If all the new points are valid and guarantee feasibility, the move is added to the move list. The whole process is enumerated until the $next(generator)$ function produces the last generator string. The procedure $skip(generator,p)$ allows necessary pruning in the segment search.

\subsection{Heuristics}

Local search algorithms guided by a single heuristic function often gets trapped in plateaus or local minima. One heuristic can possibly take the search out of the trap of local minima of another heuristic. In stead of guiding the search by a single fitness function, we maintain three different heuristics and select one of them at each iteration. We explored a number of heuristics in our experiments. The best three are finally used:
\begin{enumerate}

\item {\textbf{Maximize pairwise H-H contacts:}}
Select a move that minimizes the number of contacts between two non-consecutive amino-acids. 
\begin{equation*}
h_1=\sum_{i+1<j, s_i=H,s_j=H}^{n} c_{ij} %\times (s_i=H,s_j=H)
\end{equation*}
Here, $c_{ij}$ = 1 only if two monomers $i$ and $j$ are neighbors (or in contact) on the lattice and 0 otherwise. This heuristic corresponds to the HP energy function.

\item {\textbf{Minimize all pair H-H distance:}}
Select the move that minimizes the sum of the squared distances between all pairs of non-consecutive hydrophobic amino-acids. The heuristic is defined below:
\begin{equation*}
h_2=\sum_{i+1<j, s_i=H,s_j=H}^{n} d(i,j)^2 % \times (s_i=H,s_j=H)
\end{equation*}
Here, $d(i,j)$ denotes the Euclidean distance between the positions of $i$ and $j$ monomers. This fitness function helps pull all the hydrophobic residues towards each other and form a compact core quickly. 

\item {\textbf {Minimize squared distance to hydrophobic centroid:}}
Select the move that minimizes the sum of distances of the H-amino acids to the hydrophobic centroid ($H_c$). We calculate the co-ordinates of the hydrophobic centroid from the average of Cartesian co-ordinates of the hydrophobic amino-acids.
\begin{equation*}
 x_c=\frac{1}{n_H}\sum_{i_H=0}^{n_H} x_{i_H}, y_c=\frac{1}{n_H}\sum_{i_H=0}^{n_H} y_{i_H}, z_c=\frac{1}{n_H}\sum_{i_H=0}^{n_H} z_{i_H} 
\label{eqHCC}
\end{equation*}
Now the sum of the distances to this hydrophobic centroid ($H_c$) is defined below:
\begin{equation*}
h_3=\sum_{i_H=0}^{n_H} (x_c-x_{i_H})^2+(y_c-y_{i_H})^2+(z_c-z_{i_H})^2
\end{equation*}
Here, $n_H$ is the number of hydrophobic amino-acids in the sequence and $i_H$ is the index of an amino acid in the sequence. 
\end{enumerate}
 
We also explored several other heuristics: minimizing distance from the centroid, where the centroid is defined as $(\frac{1}{n}\sum_{i=1}^{n}x_i, \frac{1}{n}\sum_{i=1}^{n}y_i,\frac{1}{n}\sum_{i=1}^{n}z_i)$; minimizing the distance from the origin defined as $\sum_{i=1}^{n}(x_i^2+ y_i^2+z_i^2$; and maximizing the sum of neighboring contacts for hydrophobic residues defined as $\sum_{i_H}^{n_H}neighbors(i_H)$. However, these heuristics were not effective and not chosen for the algorithm.

\begin{table*}[t]
\caption{Energy levels for the R, f180 and CASP instances\label{tableMain}}
%\begin{scriptsize}
\renewcommand{\arraystretch}{1.4}
\begin{center}
\begin{tabular}{c|c|c||cc||ccc||ccc||c}

\hline
\cline{1-12}
{}&{}&{}&\multicolumn{2}{c||}{\bf LWS}&\multicolumn{3}{c||}{\bf LS-Mem}&\multicolumn{3}{c||}{\bf LS-Tabu}&{}\\
\cline{4-11}
{Seq.}&{Len}&{$E_{l}$}&{best}&{avg}&{best}&{avg}&{R.I.\%}&{best}&{avg}&{R.I.\%}&{\bf LNS}\\
\hline
\cline{1-12}
{R1}&{200}&{-384}&{\textit{-359}}&{\bf -346}&{-353}&{-326}&{34.48}&{-332}&{-318}&{42.42}&{-330}\\
{R2}&{200}&{-383}&{\textit{-360}}&{\bf -346}&{-351}&{-330}&{30.18}&{-337}&{-324}&{37.28}&{-333}\\
{R3}&{200}&{-385}&{\textit{-356}}&{\bf -349}&{-352}&{-330}&{34.54}&{-339}&{-323}&{41.93}&{-334}\\
{f180$_1$}&{90}&{-378$^*$}&{\textit{-362}}&{\bf -346}&{-360}&{-334}&{27.27}&{-338}&{-327}&{37.25}&{-293}\\
{f180$_2$}&{90}&{-381$^*$}&{\textit{-365}}&{\bf -354}&{-362}&{-340}&{34.14}&{-345}&{-334}&{42.55}&{-312}\\
{f180$_3$}&{90}&{-378}&{\textit{-367}}&{\bf -356}&{-357}&{-343}&{37.14}&{-352}&{-339}&{43.58}&{-313}\\
\hline
{3no6}&{229}&{-455}&{\textit{-416}}&{\bf -397}&{-400}&{-375}&{27.5}&{-390}&{-373}&{29.26}&{-}\\
{3mr7}&{189}&{-355}&{\textit{-320}}&{\bf -305}&{-311}&{-292}&{20.63}&{-301}&{-287}&{26.47}&{-}\\
{3mse}&{179}&{-323}&{\textit{-285}}&{\bf -270}&{-278}&{-254}&{23.18}&{-266}&{-249}&{28.37}&{-}\\
{3mqz}&{215}&{-474}&{\textit{-422}}&{\bf -408}&{-415}&{-386}&{25}&{-401}&{-383}&{27.47}&{-}\\
{3on7}&{279}&{?}&{\textit{-509}}&{\bf -493}&{-499}&{-463}&{-}&{-491}&{-461}&{-}&{-}\\
{3no3}&{258}&{-494}&{\textit{-414}}&{\bf -394}&{-397}&{-361}&{24.81}&{-388}&{-359}&{25.92}&{-}\\
\hline
\cline{1-12}
\end{tabular}
\end{center}
%\end{scriptsize}
\end{table*}

\subsection{Implementation}
We have implemented our algorithm in C++. Cartesian co-ordinates of the amino acids are used in representing protein structures and only the feasible structures are allowed. The representation and the search process ensure the satisfaction of the constraints. The performance of the local segment search mainly depends on the move generation and heuristics calculation at each iteration. Moves are generated by the iterative procedure $generateMoves$. The heuristics are maintained using the invariants provided by Kangaroo \cite{NewtonPSM11} which is a constraint based local search (CBLS) system. Invariants are used in defining mathematical operators over the variables. Calculations due to simulation and execution are performed incrementally by Kangaroo.

\section{{Experimental Results}}
\label{secRes}
We ran experiments on a cluster machine. The cluster has a number of machines each equipped with two 6-core CPUs (AMD Opteron @2.8GHz, 3MB L2/6M L3 Cache) and 64GB Memory, running Rocks OS. We compared the performance of our algorithm with the tabu search \cite{dotu2011protein} and the memory based search \cite{mem2012}\footnote{Source code were provided by the authors}. Throughout this section, tabu search and the memory-based search are denoted by LS-Tabu (or LS-T) and LS-Mem (or LS-M). For each of the protein sequences, we ran each algorithm for 50 times given 5 hours time cutoff. Our algorithm, denoted by LWS was initialized by the best solutions found by LS-Mem in 20 minutes.

\paragraph{Benchmark Set - I.}
The first benchmark set is taken from Sebastian Will's PhD thesis \cite{will2005phd}. These are the $R$ and $f180$ sequences of length 200 and 180 respectively. The best and average energy levels achieved are reported in the upper part of Table~\ref{tableMain}. Parameter settings for LS-Tabu and LS-Mem were set as suggested by the authors. We ran our algorithm with $segmentSize$ and $maxStable$ initially set to 1 and 1000 respectively, and multiplying $factor$ was set to 1.2. We could not run the large neighborhood search algorithm in \cite{dotu2011protein} on these benchmarks since the COMET program exited with `too much memory needed' on our system. However, the best energy levels from their paper are shown in the `LNS' column. Optimal lower bounds of the minimum energy values for the proteins are also reported under the column ‘$E_l$ ’ generated by the CPSP tools \cite{mann2008cpsp}. Note that these values are obtained by using exhaustive search methods and are only used to see how far our results are from them. The missing values indicate where no such bound was found and the values marked with * means the algorithm did not converge.

\paragraph{Benchmark Set - II.} 
The second set of benchmarks, derived from the famous CASP competition\footnote{http://predictioncenter.org/casp9/targetlist.cgi}, were originally used in \cite{mem2012}. Six proteins randomly chosen from the target list with length around $230\pm50$ are converted to HP sequences depending on the nature of the amino acids. PDB ids and results for these six proteins are also reported in Table~\ref{tableMain} (lower part). LNS column contains no data for these six proteins since they were not used in \cite{dotu2011protein}.

\paragraph{Tortilla Benchmarks.}
Tortilla benchmarks or ``Harvard'' benchmarks have been extensively used in the literature. All these proteins are 48 in size. We do not report the best or average energy levels for these sequences, since all of the three algorithms reach near optimal results and the difference is very small in terms of energy level. Instead, for each algorithm, we report in Table~\ref{tableTortilla} the success rates to reach the optimal structures. Time cutoff was 10 minutes for these small proteins.
    
\begin{table*}[!htb]
\begin{center}
%\setlength{\tabcolsep}{1pt}
%\begin{scriptsize}
\renewcommand{\arraystretch}{1.4}
\begin{tabular}{c|c|c|c|c|c||c|c|c|c|c|c}
\hline
\cline{1-12}
{}&{}&\multicolumn{4}{|c||}{\bf Success Rate (\%)}&{}&{}&\multicolumn{4}{c}{\bf Success Rate (\%)}\\
\cline{3-6}
\cline{9-12}
{Seq}&{$E_{l}$}&{LWS}&{LNS}&{LS-M}&{LS-T}&{Seq}&{$E_{l}$}&{LWS}&{LNS}&{LS-M}&{LS-T}\\
%\cline{1-8}
\hline
\cline{1-12}
{H1}&{-69}&{\bf 32}&{6}&{4}&{2}&{H6}&{-70}&{\bf 16}&{0}&{0}&{0}\\
%\hline 
{H2}&{-69}&{\bf 18}&{4}&{2}&{2}&{H7}&{-70}&{\bf 12}&{0}&{0}&{0}\\
%\hline
{H3}&{-72}&{\bf 24}&{0}&{0}&{0}&{H8}&{-69}&{\bf 20}&{4}&{2}&{2}\\
%\hline
{H4}&{-71}&{\bf 26}&{10}&{0}&{0}&{H9}&{-71}&{\bf 16}&{4}&{0}&{0}\\
%\hline
{H5}&{-70}&{\bf 22}&{10}&{2}&{0}&{H10}&{-68}&{\bf 24}&{8}&{2}&{0}\\
%\hline
\hline
\cline{1-12}
\end{tabular}
%\end{scriptsize}
\caption{Success rates for Torilla benchmarks, LS-Tabu and LS-Mem are denoted by LS-T and LS-M %of LWS compared to LNS 
\label{tableTortilla}}
\end{center}
\end{table*}

\subsection{Analysis}
From the average energy levels shown in bold-faced fonts in Table~\ref{tableMain} and the success rates shown in Table~\ref{tableTortilla}, it is clearly evident that our algorithm performs significantly better than the state-of-the-art algorithms. We also report new lowest energy levels for all 12 proteins in the italic fonts shown in Table~\ref{tableMain}. The success rates to reach the optimal energy levels for the Tortilla benchmarks are also higher for our algorithm as shown in Table~\ref{tableTortilla}.

\paragraph{Relative Improvement.}
We report the relative achievement of our approach measured in terms of the difference with optimal bound of the energy level in the `R.I.\%' column of Table~\ref{tableMain}. This value is significant because it gets harder to find better conformations as the energy level of a protein sequence approaches the optimal. Similar measurements are also used in \cite{mem2012}. Relative improvement (R.I.) is defined as:
%\begin{small}
\begin{equation*}
  R. I. = \frac{E_o - E_r} {E_{l} - E_r} \times 100\%
  \label{eqRE}
\end{equation*}
%\end{small} 
  where $E_o$ is the average energy level achieved by our approach, $E_r$ is the average energy level achieved by the other approach, and $E_{l}$ is the optimal lower bound of the energy level. The missing values indicate the absence of any lower bound for the corresponding protein sequence. For all the proteins, our method achieves significant improvement; which we confirmed by performing \textit{t}-test with 95\% confidence level.

\begin{figure}[t]
\begin{center}
\includegraphics[width=0.5\textwidth]{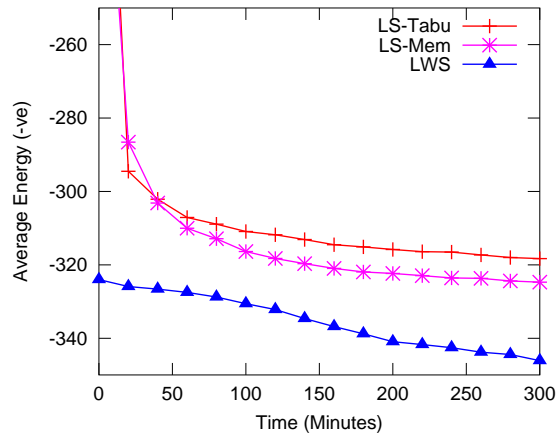}
\end{center}
\caption{Search progress for protein sequence R1 %with time for three algorithms 
\label{figProgress}}
\end{figure}

\paragraph{Search Progress.}
We show search progress of three algorithms for the protein sequence R1 in Figure~\ref{figProgress}. Average energy levels achieved by each of the algorithms for 50 runs are shown. LS-Tabu and LS-Mem achieve almost the same levels of energy initially, but as soon as the search makes progress, they fail to overcome stagnation and do not improve after a certain level. However, LWS starts from a low energy level and keeps improving the solutions. It adjusts the $segmentSize$ dynamically, which results in more perturbation and produces better results.

%\begin{figure}[!htb]
%  \begin{center}
%    \begin{tabular}{cc}
%          \includegraphics[width=.19\textwidth]{./tabu.eps}
%      &
%          \includegraphics[width=.199\textwidth]{./our.eps}
%      \\
%      (a)&(b)\\
%    \end{tabular}
%  \end{center}
%  \caption{\footnotesize Structures produced by (a) LS-Mem and (b) LWS}%
%  \label{figstruct}%
%\end{figure}

\section{{Conclusion}}
\label{secCon}
In this paper, we have presented a hybrid local search that exhaustively explores segments of a conformation and is guided by a mix of heuristic functions. Our algorithm  improved on standard benchmark proteins and significantly outperformed state-of-the-art algorithms. We applied single large segments and multiple small segments, and dynamically adjusted the segment size with stagnation period. We used several heuristics so that one heuristic can help escape local minima of another. In future, we wish to apply these techniques in other domains such as, satisfiability and traveling salesman problem.

\bibliographystyle{unsrt}
{\small
\bibliography{cambio13}}

\end{document}